\documentclass[preprint,12pt]{elsarticle}

\usepackage{amssymb}

\usepackage{amsmath}
\usepackage{bm}  

\journal{Journal of Non-Crystalline Solids}

\begin{document}

\begin{frontmatter}



\title{Determination of the Electronic Localized-to-Extended-State Transition Point around the Metal-to-Insulator Transition Region of Fluid Mercury Using the Framework of Multifractal Analysis}


\author[label1]{Kentaro Kobayashi}
\author[label1]{Takuya Sekikawa}
\author[label2]{Kenji Maruyama}

\address[label1]{Graduate School of Science and Technology Niigata University, Niigata, 950-2181, Japan}
\address[label2]{Faculty of Science Niigata University, Niigata, 950-2181, Japan}

\begin{abstract}
The metal-to-insulator transition and the presence of the disorder induced localization of electronic orbitals of fluid mercury (f-Hg) were investigated. The electronic structure of f-Hg was simulated by means of \textit{ab initio} molecular dynamics. To see the behavior of the simulation size dependency of the electronic orbital, systems with different number of atoms were simulated. The size dependency of the multifractal measure of the electronic orbitals reversed with decreasing the density, which suggests there exists the extended-to-localized-state transition near the metal-to-insulator transition region of f-Hg.
\end{abstract}


\end{frontmatter}



\section{Introduction}

Several experimental measurements showed that fluid mercury (f-Hg) undergoes metal-to-insulator transition (MIT) during the expansion along the liquid-gas coexistence curve and over the liquid-gas critical point in the $P$-$T$ diagram. For example, electrical conductivity measurements \cite{1966_Kikoin_conductivity,1968_Hensel_conductivity,1979_Schonherr_conductivity,1982_Yao_thermo,1990_Hensel_conductivity} showed that electrical conductivity of f-Hg continuously decreases as the density decreases and it crosses the value of $10^2$ ${\rm \Omega}$ cm$^{-1}$ at around the density of 8 g cm$^{-3}$. This density of 8 g cm$^{-3}$ is the approximate value of the MIT, where the knight shift disappears \cite{1975_El-Hanany_Knight,1982_Warren_Knight} and the optical gap appears \cite{1990_Yao_optical,1993_Yao_optical,1996_Hensel_optical}.

As of a usual explanation, the mechanism of MIT of f-Hg is basically explained as the Bloch-Wilson band crossing transition \cite{1976_Yonezawa_TightBinding}. In addition to the band crossing transition, the existence of an effect of atomic disorder of the f-Hg was suggested. If the interatomic potential is highly disordered, electrons are localized and the system becomes insulator even if there exists finite density of states (DOS) at the Fermi energy ($E_{\rm F}$), the scheme of which was proposed by Mott \cite{1966_Mott_Hg}. Mott also proposed that localized states near $E_{\rm F}$ and extended states away from $E_{\rm F}$ are separated by the mobility edge.
Cohen and Jortner \cite{1973_Cohen_percolation,1974_Cohen_percoltion} described metal-nonmetal transition of f-Hg using percolation theory with the relation to the heterogeneity of f-Hg under the high temperature.
Winn and Logan \cite{1990_Winn_localization} studied this mechanism of MIT of f-Hg, especially on the relation between the extended-to-localized-state transition and the Bloch-Wilson transition of electron band.
Bruce and Morgan \cite{1995_Bruce_QI} paid attention to the quasielastic scattering of electrons and showed that the change of the conductivity by quantum interference for the liquid metals was comparable to that for the amorphous alloys.

Kresse \textit{et al.} \cite{1996_Kresse_AbInitio_early,1997_Kresse_AbInitio} performed \textit{ab initio} molecular dynamics (MD) of f-Hg with 50 atoms. They calculated the participation ratio \cite{1970_Bell_ParticipationRatio,1972_Edwards_ParticipationRatio} to evaluate the localization of the electronic orbitals. Their results showed that the DOS at $E_{\rm F}$ of f-Hg decreased with the decreasing of the density and it became zero at the density between 8.78 and 5.8 g cm$^{-3}$, whereas the participation ratio near the $E_{\rm F}$ showed weak trend to lower value at the density of 5.8 g cm$^{-3}$. They concluded that MIT of f-Hg is explained mainly with the Bloch-Wilson transition and the extended-to-localized-state transition playing only the minor role. Carder\'in \textit{et al.} \cite{2011_Calderin_AbInitio_early,2011_Calderin_AbInitio} also performed \textit{ab initio} MD of f-Hg with 90 atoms. Their results showed similar DOS's with those obtained by Kresse \textit{et al.} \cite{1997_Kresse_AbInitio}. Their results also showed that the DOS at $E_{\rm F}$ of f-Hg becomes zero near the density of 7.0 g cm$^{-3}$.
 Unfortunately, as the participation ratio continuously varies within the finite value, one can't determine the electronic extended-to-localized-state-transition point clearly.

We focus on following two methodologies for the determination of this transition point. 
The first methodology is the scaling theory by Abrahams \textit{et al.} \cite{1979_Abrahams_scale,1981_MacKinnon_transmission}.
According to Abrahams \textit{et al.} \cite{1979_Abrahams_scale}, in the case of three dimension, there exists an unstable critical point between localized and extended systems. Also, the conductance of the system, that is not at the critical point, increases or decreases monotonically with the change of the scale of the system. These property imply that most systems are classified as localized or extended state when the size of the system is infinite. 
The numerical studies on the scaling behavior of the electronic orbital under the uniform random potential distributed with a width of $W$ were performed \cite{1981_Pichard_transferMatrix,1981_MacKinnon_transmission,1983_MacKinnon_transmission2,2018_Slevin_CriticalExponent}. MacKinnon \textit{et al.} calculated the localization length of quasi one dimensional systems and obtained the critical disorder $W_{\rm c}$ for three dimensional case as about 16.5. Moreover, recent study \cite{2018_Slevin_CriticalExponent} using the transfer matrix method reported the value of $W_{\rm c}$ about 16.543.

The second methodology is the multifractal analysis of electronic orbital \cite{2008_Vasquez_MFA,2008_Rodriguez_MFA,2010_Rodriguez_MFA,2019_Carnio_MFA}.
Rodriguez \textit{et al.} \cite{2010_Rodriguez_MFA} showed that the scale invariant feature of electronic orbitals at the criticality can be described using the framework of the multifractal analysis. They calculated the multifractal measure $\alpha$ defined by the following equation:
\begin{equation}
\label{eq1}
  \alpha = \log \mu / \log \lambda \quad ,
\end{equation}
where $\mu= \int_B {|\psi|^2} d \boldsymbol{r}$, $\psi$ is the orbital, $\boldsymbol{r}$ is the coordination of real space, $B$ is the space of a small box and $\lambda$ is the ratio of linear size of the box $B$ to linear size of original cell. With the $W$ much larger than $W_{\rm c}$, \textit{e.g.} $W$=18, $\alpha$ with fixed $\lambda$ increased with increasing of the system size. On the other hand, with $W$ much smaller than $W_{\rm c}$, \textit{e.g.} $W$=15, it decreased with increasing of system size. At $W$ near the critical point, $\alpha$ showed scale invariant feature. They also considered the finite size scaling feature and derived the value of the critical disorder $W_{\rm c}$ as 16.56. 
Though the relation between the transfer matrix method and the multifractal analysis is not obvious, both method gave the very close value of $W_{\rm c}$. We relied on this fact and used the framework of multifractal analysis to determine the criticality of extended-to-localized-state transition of f-Hg.

In the recent years, the multifractal analyses toward electronic state began to be applied to more figurative objects, for example, analysis of sulfur-doped silicon using tight-binding model constructed with \textit{ab initio} calculation \cite{2019_Carnio_ETB,2019_Carnio_MFA}, quasicrystals \cite{2017_Mace_quasicrystal} and dilute magnetic semiconductor using scanning tunneling microscopy \cite{2010_Richardella_STM}. These works used the multifractal analysis aiming to clarify the critical behavior, which is displayed as the multifractal spectrum \cite{2002_Riedi_MP}.

\section{Method}

The \textit{ab initio} MD simulations for f-Hg were performed using Quickstep scheme implemented in the CP2K program package \cite{2005_Quickstep} with Perdew-Burke-Ernzerhof functionals based on the GGA \cite{1996_PBE_GGA}. The Kohn-Sham orbitals were expanded in Gaussian basis set with double-$\zeta$ plus polarization quality, whereas the electron density was expanded in plane waves up to a cutoff of 400 Ry. Goedecker-Teter-Hutter \cite{1996_GTH_pseudopotential} (GTH) pseudopotential with 12 valence electrons ($5d^{10} 6s^{2}$) was used for Hg. Brillouin zone integration was restricted to the $\Gamma$ point. Target convergence threshold of $10^{-6}$ Hartree was used for the self-consistent field (SCF) calculation.

The initial atomic configurations were created by a set of reverse Monte Carlo simulations with X-ray diffraction data \cite{1998_Tamura_XRay}. 
The linear sizes of systems for every density as well as the number of atoms are listed in the table \ref{linearsize}.
 The \textit{ab initio} MD simulations in the $NVT$ ensemble were carried out using the GLE thermostat \cite{2009_Ceriotti_GLEThermostat2,2009_Ceriotti_GLEThermostat}. Time step for the system with temperature of 293.15 K was set as 6 fs. For the other temperatures, time steps were scaled so as the mean displacement of atoms per one MD step were to be the same with that for 293.15 K. The thermalizations were done in the first 1000 steps. The samplings were done in the next 2000 steps for the system of 100 atoms and 1000 steps for the systems of 200 atoms and 400 atoms.

The multifractal measure $\alpha$ was calculated using the Eq.\ref{eq1} with the fixed $\lambda$ of value 0.50. To obtain the distribution of $\alpha$ in a MD cell, the coordination of the small box $B$ was placed at every point of the space of the MD cell as:
 \begin{equation}
\label{eq2}
  \mu(\boldsymbol{r}) = ({\rm rect} \ast |\psi|^2)(\boldsymbol{r})  \quad,
\end{equation}
which is the periodic convolution of ${\rm rect}(\boldsymbol{r})$ and $|\psi(\boldsymbol{r})|^2 $, where ${\rm rect}(\boldsymbol{r})$ is a support of $B$ or a rectangular filter function, $\mu(\boldsymbol{r})$ is $\mu$ for $B$ translated with vector $\boldsymbol{r}$. Then $\alpha(\boldsymbol{r})$ is obtained using Eq.\ref{eq1} as: $  \alpha(\boldsymbol{r}) = \log \mu(\boldsymbol{r}) / \log \lambda $ . The mean of $\alpha$ for a single orbital is defined as:
 \begin{equation}
\label{eq3}
  \alpha_{\rm m}= \frac{1}{V} \int_{\rm MD\ cell} { \alpha(\boldsymbol{r}) } d \boldsymbol{r} \quad ,
\end{equation}
where $V$ is the volume of the MD cell. 
For the systems with 400 atoms or 200 atoms, $\alpha$ was sampled from 11 different configurations of atoms of the intervals of 100 MD steps. For 100 atom system, $\alpha$ was sampled from 21 different configurations of the intervals of 100 MD steps. These numbers are listed in Table \ref{num_md}. As the result of statistics of the different orbitals in the certain energy range and the atomic configurations, we can consider $\langle\alpha_{\rm m}\rangle$ and $s_{\alpha_{\rm m}}$, which are the mean value and the standard deviation of $\alpha_{\rm m}$, respectively. Since the order of the averaging operations for the origin of box and for the orbitals and atomic configurations are commutative, the $\langle\alpha_{\rm m}\rangle$ could be ambiguously called as the mean value of $\alpha$, but not for the $s_{\alpha_{\rm m}}$.

\section{Results}

\begin{figure}[ht]%
  \includegraphics*[width=8.5cm]{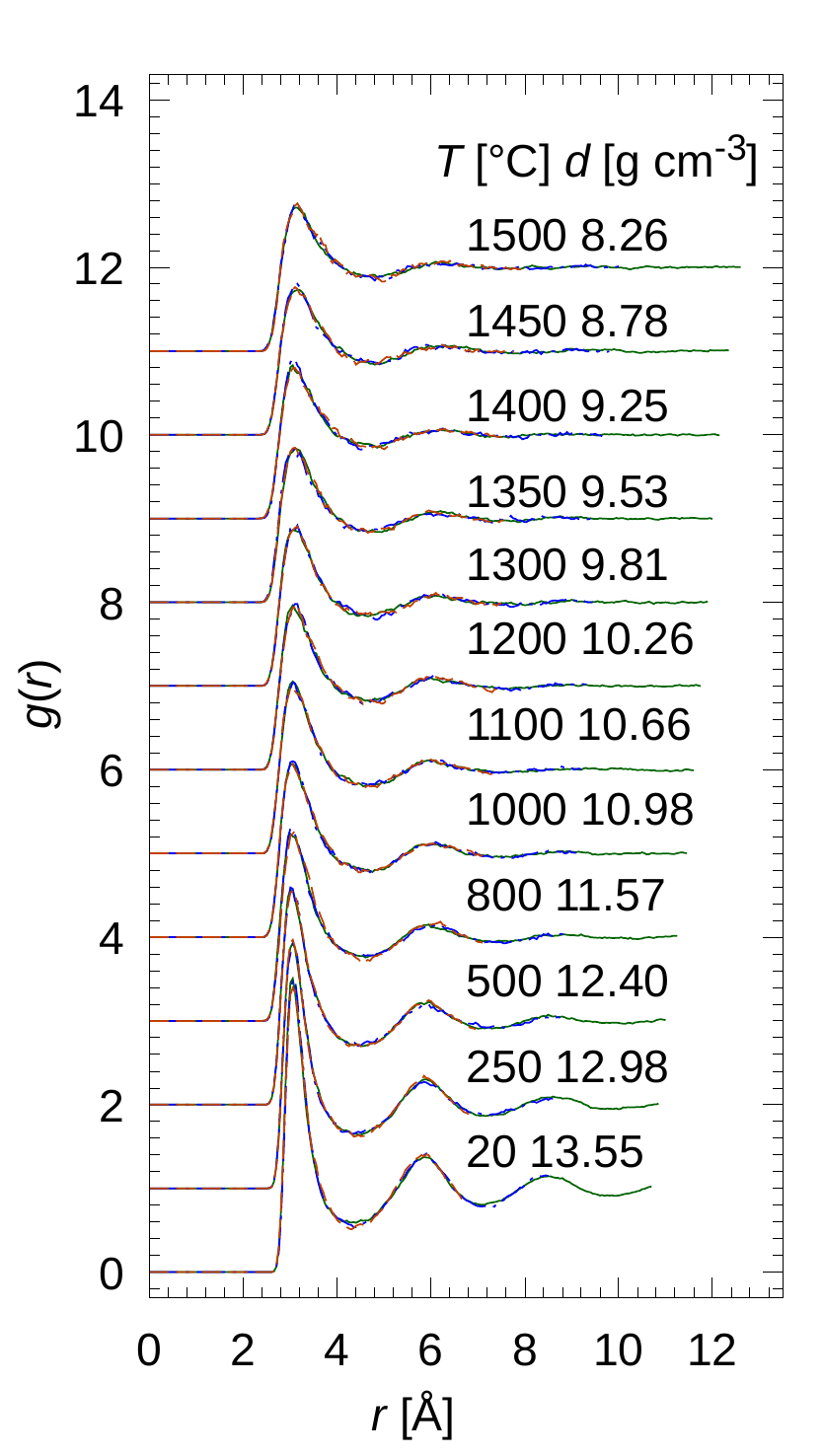}%
  \caption{The pair distribution function $g(r)$ of f-Hg at the densities from 13.55 to 8.26 g cm$^{-3}$. The successive curves are displaced by 1.0 for clarity. The green solid, the blue dot-dashed and the red dashed curves represent for the system of 400, 200 and 100 atoms, respectively.}
    \label{gr}
\end{figure}

The pair distribution functions $g(r)$ of atomic structures of f-Hg are shown in Fig.\ref{gr}. For the systems of 100, 200 and 400 atoms with the same temperatures and densities, $g(r)$'s were in accord except for the maximum of $r$ which is defined by half of the linear size of MD cell. These $g(r)$'s obtained in this MD study almost agreed with the X-ray diffraction data \cite{1998_Tamura_XRay,2003_Inui_XRay}. Note that for the density of 10.66 g cm$^{-3}$, the  shape of the first peak agreed better with the data of 10.7 g cm$^{-3}$ by Inui \textit{et al.} \cite{2003_Inui_XRay} rather than the earlier data by Tamura \textit{et al.} \cite{1998_Tamura_XRay}. The data with the densities smaller than about 10.98 g cm$^{-3}$ also showed good agreement with the data by Inui \textit{et al.}

\begin{figure}[ht]%
  \includegraphics*[width=8.5cm]{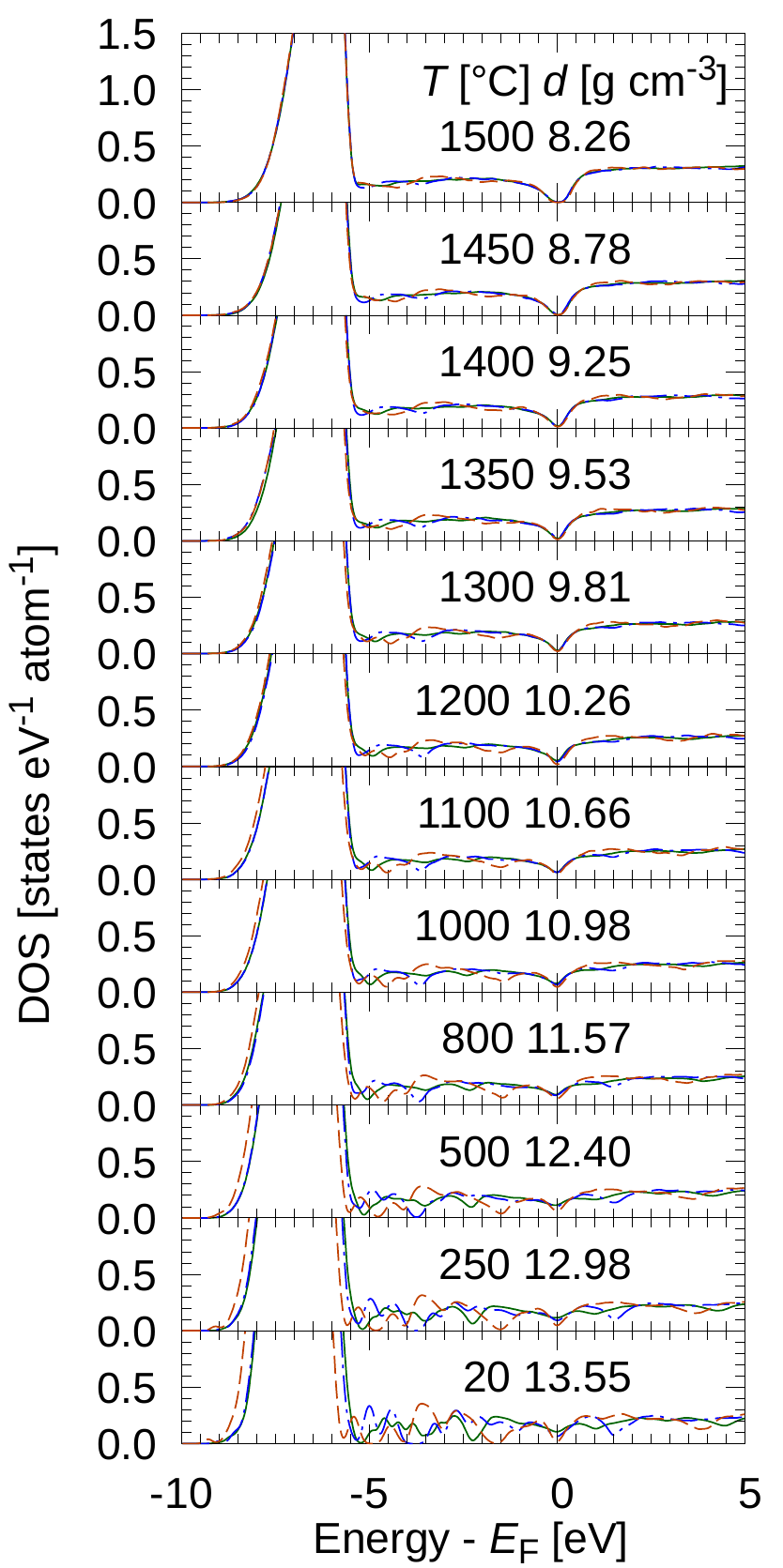}%
  \caption{The density of states of f-Hg as a function of energy. The green solid, the blue dot-dashed and the red dashed curves represent for the system of 400, 200 and 100 atoms, respectively.}
    \label{dos}
\end{figure}

Fig.\ref{dos} shows the electronic density of states (DOS) of f-Hg. The DOS plotted here was broadened with the normal distribution function with the standard deviation of 0.1 eV. The DOS at $E_{\rm F}$ decreased with decreasing the density, became almost zero at the density of 8.78 g cm$^{-3}$. For the DOS of the density of 13.55 g cm$^{-3}$, a pseudogap appeared for the system of 100 atoms; however, it was filled up as the numbers of atoms increased.

\begin{figure}[ht]%
  \includegraphics*[width=7.5cm]{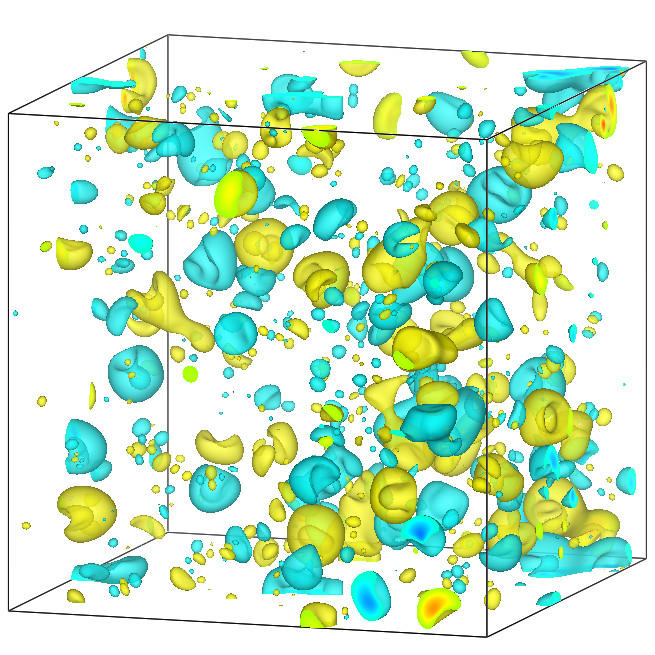}%
  \caption{The contour of the orbital of f-Hg for the density of 12.40 g cm$^{-3}$ at around $-0.2$ eV from $E_{\rm F}$ for systems of 400 atoms. Iso-density value of $ |\psi|^2 = 6.4 \times 10^{-5}$ bohr$^{-3}$ was plotted. The blue and yellow colors distinguish the phase of $\psi$.}
    \label{o3}
\end{figure}

\begin{figure}[ht]%
  \includegraphics*[width=7.5cm]{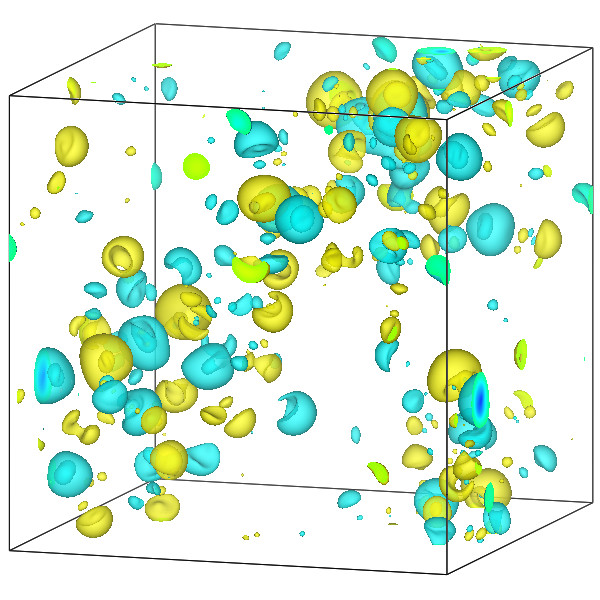}%
  \caption{The contour of the orbital of f-Hg for the density of 8.26 g cm$^{-3}$ at around $-0.2$ eV from $E_{\rm F}$ for systems of 400 atoms. Iso-density value of $ |\psi|^2 = 6.4 \times 10^{-5}$ bohr$^{-3}$ was plotted. The blue and yellow colors distinguish the phase of $\psi$.}
    \label{o12}
\end{figure}

The contours of wavefunction of the electronic orbital of f-Hg are shown in Figs.\ref{o3} and \ref{o12}. The set of multifractal measures $\alpha$ were calculated for each orbital.

\section{Discussion}

\begin{figure}[ht]%
  \includegraphics*[width=8.5cm]{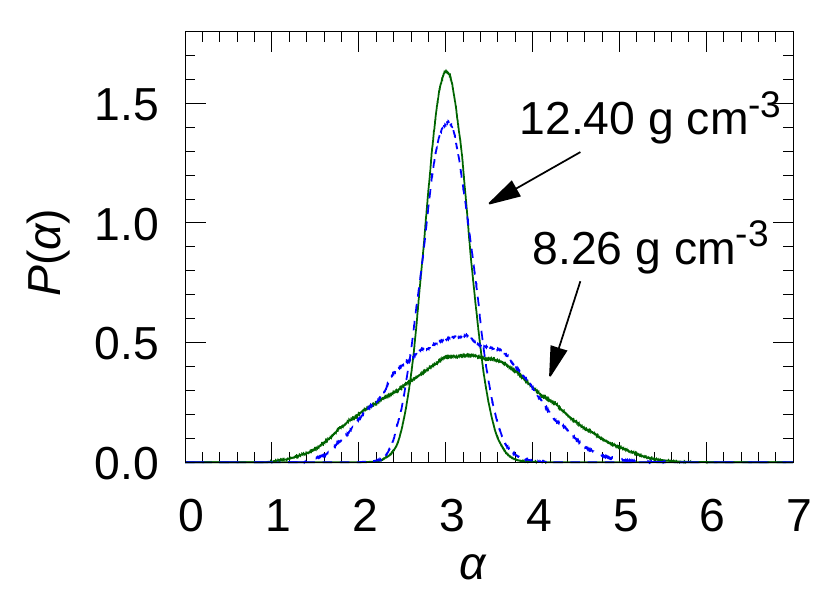}%
  \caption{The distribution of the multifractal measure $\alpha$. Curves are for the densities of 12.40 and 8.26 g cm$^{-3}$. The green solid, the blue dashed curves represent the $P(\alpha)$ for the system of 400 and 200 atoms, respectively. The energy levels of the orbitals for the $P(\alpha)$ shown in this figure are in the region of $-0.3$ to $-0.1$ eV from $E_{\rm F}$.}
    \label{distribution}
\end{figure}

The distributions of multifractal measure $\alpha$: $P(\alpha)$ sampled for the energy levels from $-0.3$ to $-0.1$ eV are shown in Fig.\ref{distribution}. The $P(\alpha)$ for the density of 12.40 g cm$^{-3}$, which correspond to the metallic state region, became narrower as the number of atoms increased from 200 to 400. On the other hand, the $P(\alpha)$ of the density of 8.26 g cm$^{-3}$ became wider as the number of atoms increased. This feature could be corresponded with the report of the electronic orbitals under uniform random potential by Rodriguez \textit{et al.} \cite{2010_Rodriguez_MFA}. Their result shows the $P(\alpha)$ becomes narrower as the system size increases when disorder parameter $W$ is 15 which is much smaller than the extended-to-localized-state transition disorder $W_{\rm c}$ ($\approx$16.543) \cite{2018_Slevin_CriticalExponent}. Also, when $W$=18, which is much larger value, $P(\alpha)$ becomes wider as the system size increases. These two features show the possibility of f-Hg undergoing the extended-to-localized-state transition near MIT. Also, it suggests that for f-Hg, the potential disorder increase monotonically as the density decrease.

\begin{figure}[ht]%
  \includegraphics*[width=8.5cm]{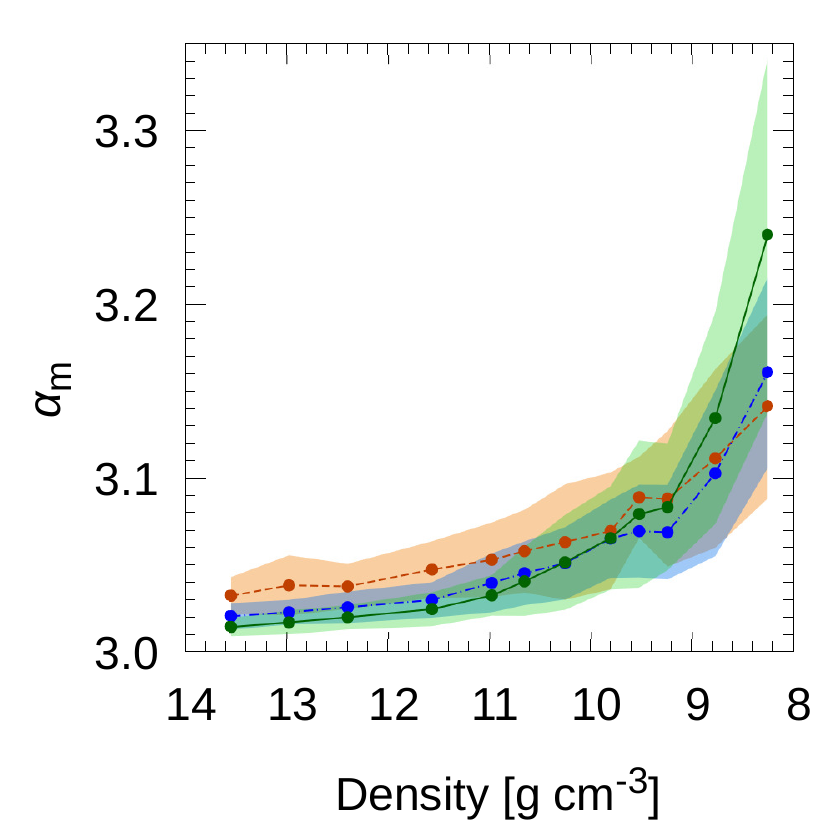}%
  \caption{The mean value $\alpha$ of the distribution $P(\alpha)$ of f-Hg as a function of the density. The green solid, the blue dot-dashed and the red dashed curves represent the $\alpha_{\rm m}$ for the system of 400, 200 and 100 atoms, respectively. The green, blue and red shaded areas represent the $\pm 1$ standard deviation uncertainty of $\alpha_{\rm m}$ for the system of 400, 200 and 100 atoms, respectively.}
    \label{density}
\end{figure}

Actually, Rodriguez \textit{et al.} \cite{2010_Rodriguez_MFA} analyzed the mean value of $\alpha$ rather than standard deviation of it. Therefore, we analyzed the mean value of $\alpha$, $\alpha_{\rm m}$. The $\alpha_{\rm m}$ are shown in Fig.\ref{density} as a function of density. The energy levels picked here were for $-0.3$ to $-0.1$ eV from $E_{\rm F}$.
The number of orbitals used to obtain $\alpha_{\rm m}$ at around $-0.2$ eV as a function of density is listed in the Table \ref{num_orb}. The standard deviation of $\alpha_{\rm m}$, $s_{\alpha_{\rm m}}$ is also displayed in Fig.\ref{density}.
In the Fig.\ref{density}, at the density of 12.40 g cm$^{-3}$, $\alpha_{\rm m}$ decreased as the system size increased whereas at the density of 8.26 g cm$^{-3}$, it decreased as the system size increased. As we can see in Fig.\ref{density}, the size dependency of $\alpha_{\rm m}$ reversed in the region of density between 11 and 8 g cm$^{-3}$. Further analysis would allow us figure out the finite size effect.
Rodriguez \textit{et al.} had investigated the behavior of $\alpha_{\rm m}$ with random potential of randomness $W$ \cite{2010_Rodriguez_MFA} and it can be seen from Fig. 3 of Ref.\cite{2010_Rodriguez_MFA} as follows:
A curve $A_L$ is defined as the locus on the $\alpha_{\rm m}$-$W$ plain with the system size $L$. $W_{ij}$ is $W$ value of the crossing point of curves $A_{L_i}$ and $A_{L_j}$. It can be seen that the $W_{ij}$ of finite different system sizes $L_i$, $L_j$ is larger than $W_{\rm c}$. It also can be seen that $W_{ij}$ approaches to $W_{\rm c}$ as the system sizes becomes large. Here, we associate atoms to sites in the study of Rodriguez \textit{et al.} Our results for f-Hg shows the crossing point of the curves of $\alpha_{\rm m}$ versus density of 100 atoms and 200 atoms was at about the density of 8.6 g cm$^{-3}$, whereas number of atoms of 200 and 400, 10.0 g cm$^{-3}$. We think more precise analysis required to determine the relation between this shift and the finite size effect.

\begin{figure}[ht]%
  \includegraphics*[width=8.5cm]{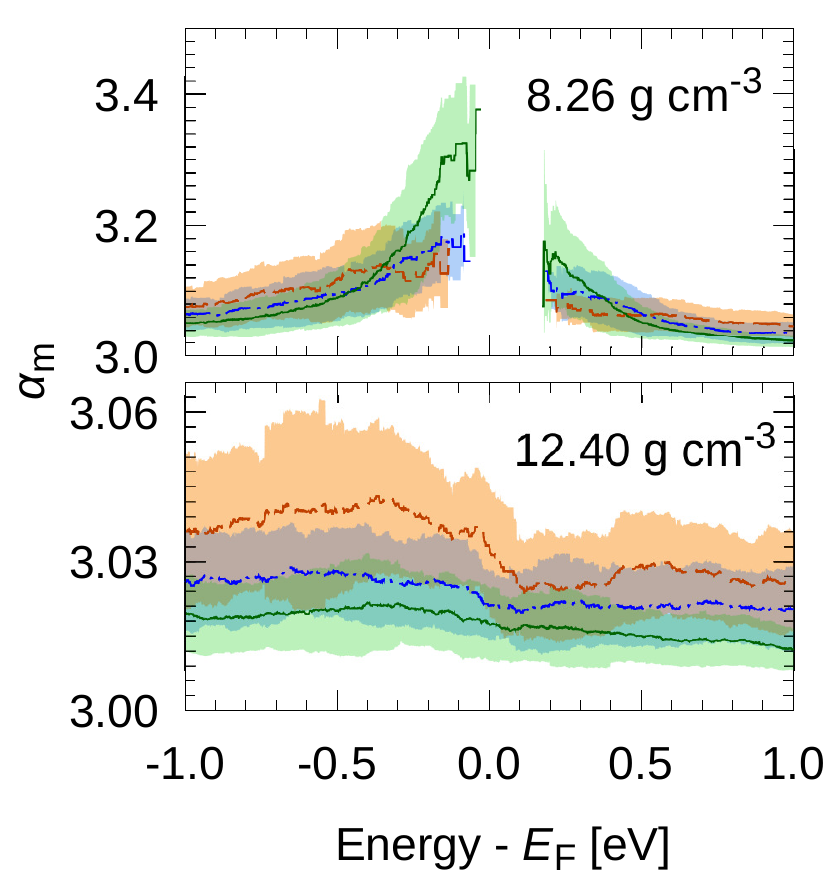}%
  \caption{The mean value of $\alpha$ as a function of energy levels of f-Hg. These curves are given by means of kernel smoothing of the set of $\alpha_{\rm m}$ using the rectangular function of width 0.2 eV. The green solid, the blue dot-dashed and the red dashed curves represent the $\alpha_{\rm m}$ for the system of 400, 200 and 100 atoms, respectively. The green, blue and red shaded areas represent the $\pm 1$ standard deviation uncertainty of $\alpha_{\rm m}$ for the system of 400, 200 and 100 atoms, respectively.}
    \label{energy}
\end{figure}

The $\alpha_{\rm m}$ as a function of energy levels are shown in Fig.\ref{energy}; in the case of Figs.\ref{distribution} and \ref{density}, only the energy level of $-0.2$ eV was shown. The $\alpha_{\rm m}$ at the density of 8.26 g cm$^{-3}$ showed discontinuous curve at the $E_{\rm F}$ due to the lack of the electronic states. For the density of 12.40 g cm$^{-3}$, at all the energy range of from $-1.0$ to $+1.0$ eV, the $\alpha_{\rm m}$ decreased as the size became large. This means that electron orbitals in this energy range were extended. At the density of 8.26 g cm$^{-3}$, the size dependency of $\alpha_{\rm m}$ was different for the energy levels. At about $-0.2$ eV, $\alpha_{\rm m}$ increased as size became larger, which means that the orbitals were localized. At about $-0.7$ eV, $\alpha_{\rm m}$ decreased as the size became larger, which means that the orbitals were extended. The energy region higher than $E_{\rm F}$ also showed that the orbitals which were close to the $E_{\rm F}$ were localized and orbitals which were far from the $E_{\rm F}$ were extended. These phenomena shows an existence of mobility edges of liquid which was proposed by Mott \cite{1966_Mott_Hg}. Further analysis would allow us determine the locus of mobility edge on the energy-density plain. Comparing with the locus of the band edges, we can clarify the relation between the band crossing transition and extended-to-localized-state transition in the MIT of f-Hg.

The reversed size dependency of $\alpha_{\rm m}$ could be observed for 100 to 400 atom systems for f-Hg, however, Carnio \textit{et al.} \cite{2019_Carnio_MFA} reported the similar phenomena for sulfur doped silicon for huge 4096 and 10648 atom system. The reason of the such difference of the system size is estimated the amplitude of the randomness of liquid material, which is much larger than that of solid materials. Therefore, similar size dependency with small number of atoms could be observed in the amorphous systems.

\section{Conclusions}

We studied fluid mercury (f-Hg) at thermodynamic states including the metal-to-insulator transition (MIT) region by means of \textit{ab initio} molecular dynamics. We simulated systems of three different number of atoms: 100, 200 and 400. The multifractal measure $\alpha$ showed opposite size dependency in the densities of 12.40 and 8.26 g cm$^{-3}$. This phenomenon suggests there exists extended-to-localized-state transition near the MIT of the f-Hg.

\section*{Acknowledgement}
One of the author (K. K.) would like to thank Yoshiaki \=Ono (Niigata University) for his advice on the Anderson localization and Masaki Saito (Niigata University) for the technical support. Numerical calculations were performed in part using Oakforest-PACS system provided by Multidisciplinary Cooperative Research Program (68) in CCS, University of Tsukuba, and MASAMUNE-IMR system provided in CCMS, IMR, Tohoku University. This work was supported by JSPS KAKENHI Grant Numbers 15K05206 and 23540441.





\bibliographystyle{model1-num-names}
\bibliography{ref------} 







\begin{table}
  \caption{The linear sizes of simulated systems in $\rm \AA$ for each density as well as the number of atoms.}
  \begin{tabular}[htbp]{@{}llll@{}}
    \hline
    Density [g cm$^{-3}$] &  100 atom & 200 atom & 400 atom \\
    \hline
     8.26 &  15.916988 & 20.054148 & 25.266643 \\
     8.78 &  15.596341 & 19.650159 & 24.757648 \\
     9.25 &  15.327582 & 19.311544 & 24.331020 \\
     9.53 &  15.175975 & 19.120530 & 24.090359 \\
     9.81 &  15.030193 & 18.936857 & 23.858945 \\
    10.26 &  14.807160 & 18.655853 & 23.504902 \\
    10.66 &  14.619589 & 18.419528 & 23.207151 \\
    10.98 &  14.476163 & 18.238822 & 22.979476 \\
    11.57 &  14.225792 & 17.923374 & 22.582037 \\
    12.40 &  13.901031 & 17.514201 & 22.066511 \\
    12.98 &  13.690816 & 17.249347 & 21.732816 \\
    13.55 &  13.496085 & 17.004001 & 21.423699 \\
    \hline
  \end{tabular}
  \label{linearsize}
\end{table}

\begin{table}
  \caption{The sampling step range of MD and the number of configurations for calculating the multifractal measure $\alpha$.}
  \begin{tabular}[htbp]{@{}llll@{}}
    \hline
     & Sampling time & Configurations \\
    \hline
    100 atom & 1000 to 3000 step & 21 \\
    200 atom & 1000 to 2000 step & 11 \\
    400 atom & 1000 to 2000 step & 11 \\

    \hline
  \end{tabular}
  \label{num_md}
\end{table}

\begin{table}
  \caption{The total number of orbitals between the energy range from $-0.3$ eV to $-0.1$ eV belonging to the MD configurations used for the calculation of the multiractal measure $\alpha_{\rm m}$. The total number of orbitals is displayed for every densities and number of atoms.}
  \begin{tabular}[htbp]{@{}llll@{}}
    \hline
    Density [g cm$^{-3}$] &  100 atom & 200 atom & 400 atom \\
    \hline
     8.26 &    6   &   14   &    28   \\
     8.78 &   17   &   26   &    41   \\
     9.25 &   25   &   26   &    66   \\
     9.53 &   35   &   38   &    64   \\
     9.81 &   35   &   39   &    79   \\
    10.26 &   37   &   42   &    78   \\
    10.66 &   44   &   49   &    86   \\
    10.98 &   44   &   48   &    88   \\
    11.57 &   41   &   46   &    98   \\
    12.40 &   45   &   56   &   107   \\
    12.98 &   53   &   57   &   106   \\
    13.55 &   47   &   47   &   111   \\
    \hline
  \end{tabular}
  \label{num_orb}
\end{table}

\end{document}